\documentclass[10pt,letterpaper]{article}
\usepackage{opex3} 

\newcommand{\ket}[1]{\left| #1\right\rangle}

\newcommand{\eeqref}[1]{Eq.~(\ref{#1})}

\usepackage{amssymb,amsmath,amsthm,amsfonts,bm,cite}
\usepackage{graphicx}
\usepackage{epsfig}
\usepackage{subfigure}
\usepackage{amsmath}			
\usepackage{color}
\usepackage{epstopdf}
\usepackage{float}

\makeatletter
\renewcommand*\env@matrix[1][c]{\hskip -\arraycolsep
  \let\@ifnextchar\new@ifnextchar
  \array{*\c@MaxMatrixCols #1}}
\makeatother

\begin{document}

\title{Direct characterization of linear-optical networks}

\author{Saleh Rahimi-Keshari,$^{1,*}$ Matthew A. Broome,$^{1,2}$ Robert Fickler,$^{3,4}$ Alessandro Fedrizzi,$^{1,2}$ Timothy C. Ralph,$^{1}$ and Andrew G. White,$^{1,2}$}
\address{$^{1}$Centre for Quantum Computation and Communication Technology, $^{2}$Centre for Engineered Quantum Systems, School of Mathematics and Physics, University of Queensland, Brisbane, QLD 4072, Australia\\
$^{3}$Quantum Optics, Quantum Nanophysics, Quantum Information,
University of Vienna, A-1090, Austria\\
$^{4}$Institute for Quantum Optics and Quantum Information, Boltzmanngasse 3, Vienna A-1090, Austria}

\email{$^{*}$s.rahimik@gmail.com} 



\begin{abstract}  
We introduce an efficient method for fully characterizing multimode linear-optical networks. Our approach requires only a standard laser source and intensity measurements to directly and uniquely determine all moduli and non-trivial phases of the matrix describing a network.
We experimentally demonstrate the characterization of a $6{\times}6$ fiber-optic network and independently verify the results via nonclassical two-photon interference.
\end{abstract}



\section{Introduction}

Implementation of quantum technologies requires the ability to realize arbitrary unitary operators, enabling applications such as efficient quantum simulation and computation. In principle, linear-optical networks, i.e., passive networks constructed from beam splitters, phase shifters and mirrors, can be used to experimentally realize any $N{\times}N$ unitary operator \cite{reck1994experimental}. A significant remaining practical challenge is to efficiently characterize the device once it is built. A known solution is to perform quantum process tomography of a device using nonclassical states \cite{Obrien:protomo,Childs2001,mitchell2003} or coherent states \cite{lobino2008,rahimi-keshari2011}. 
However, despite progress on more efficient methods such as compressive sensing \cite{PhysRevLett.106.100401}, this approach is relatively slow and impractical for large optical networks. 

A more tractable approach, starting from the assumption of linearity, would be to adapt existing methods from classical optics. As a linear-optical circuit can always be cast as an interferometer with $(N^{2}{-}N)/2$ beam splitters \cite{reck1994experimental}, it can be characterized by embedding it in an external interferometer and using a local oscillator \cite{VanWiggeran2003}. However, such a method is challenging for large networks due to the interferometric stability required. Recently, a method was proposed that obviates the use of an external interferometer~\cite{Peruzzo2011,laing2012sst}; however, it requires nonclassical interference~\cite{PhysRevLett.59.2044} for the characterization.

In this paper, we introduce an efficient method for characterizing an $N$-mode linear optical network by uniquely determining the $N{\times}N$ matrix that represents the network. It 
is an interferometric method that
uses readily available standard laser sources and photodetectors and eliminates the need for an external interferometer or nonclassical interference and single-photon detection. It is also technically simple and efficient, requiring only $2N{-}1$ configurations to directly measure all nontrivial parameters of the $N{\times}N$ matrix. We demonstrate our method by characterizing an integrated device---a $6{\times}6$ fused-fibre coupler---and highlight its precision by comparing measured quantum interference patterns with those predicted using the characterization matrix.

A linear-optical network can be represented by a linear transformation of input to output creation operators, $a^{\dagger}_{j}$ and $b^{\dagger}_{k}$ respectively, given by
\begin{equation}
a^{\dagger}_{j}=\sum_{k=1}^{N} M_{jk} b^{\dagger}_{k}.
\label{in-out-trans}
\end{equation} 
This reconstriction necessarily rules out optical amplifiers and all other non-linear elements. For an ideal lossless network the matrix $M$ is unitary. In practice however, due to loss, it is a submatrix of a larger unitary matrix. Despite this, knowledge of $M$ uniquely characterizes the device, as it determines the action of the network on any multimode coherent state $\ket{\alpha_1,\alpha_2,\ldots,\alpha_N}$, 
\begin{equation}
\beta_{k}=\sum_{j=1}^{N}M_{jk}\alpha_{j},
\label{in-out-coh}
\end{equation}  
where $\ket{\beta_1,\beta_2,\ldots,\beta_N}$ is the multimode output coherent state. Hence, following the method in~\cite{rahimi-keshari2011}, one can in principle predict the output state of the network for any given input state.

\section{Characterization Method}

Generally, the matrix elements of $M$ are complex numbers $M_{jk}{=}r_{jk} e^{i\theta_{jk}}$, where $0{\leq} r_{jk}{\leq} 1$ and $0 {\leq}\theta_{jk}{<} 2\pi$. 
Noting that the $(2N{-}1)$ phases of the basis vectors are not physically significant, we can absorb them into the basis vectors \cite{Peres}. Thus, any matrix $M$ can be decomposed as a product of three matrices 
$M{=}D(\bm{\mu}) M^{'} D(\bm{\nu})$ ,
where $D(\bm{\mu}){=}\text{diag}\left(e^{i\mu_1}, e^{i\mu_2}, \dots, e^{i\mu_N} \right)$ and both $M$ and $M^{'}$ describe the same physical network.  
Without loss of generality, we let
$\theta_{1j}{=}\theta_{j1}{=}0$, for $j{=}1,2,\dots,N$,
and we are left with $(N{-}1)^2{+}N^2$ parameters to be determined in the characterization.

Even if $M$ is unitary, it has been shown that knowledge of moduli $r_{jk}$ alone does not uniquely determine all phases for $N{>}3$ \cite{Peres,Bernstein1974}. Therefore, in order to uniquely characterize $M$ we require probe states and measurements that are sensitive to the phases $\theta_{jk}$.

\begin{figure}[b]
\centering
	\includegraphics[width=0.7\columnwidth]{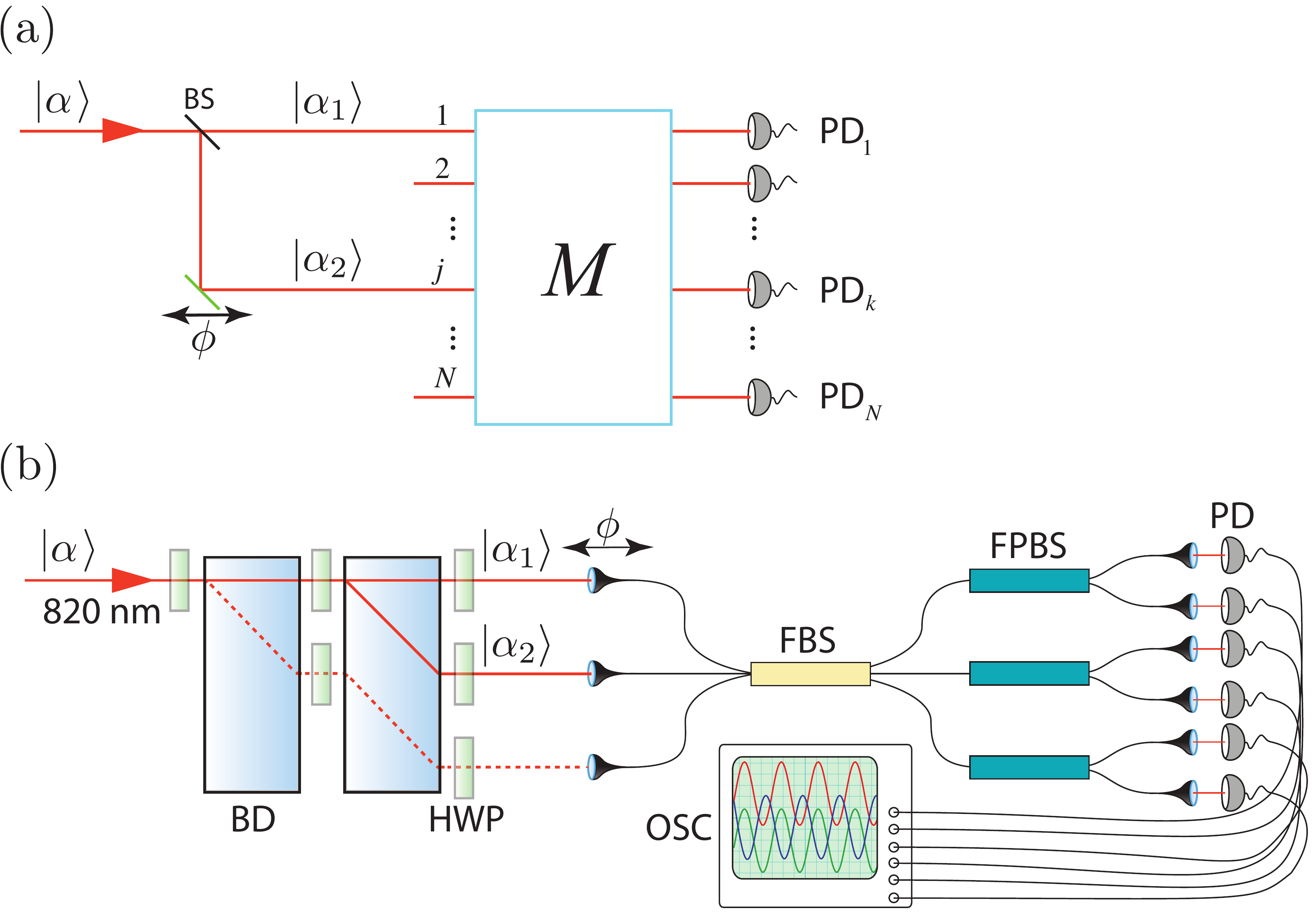}
\caption{Scheme for characterizing a linear-optical network $M$.  (a) Using a $50{:}50$ beam splitter (BS) and phase shifter ($\phi$) a dual-mode coherent state, $\ket{\alpha}$, is prepared and sent through $M$, where $\ket{\alpha_2}{=}\ket{e^{i \phi}\alpha_1}$. By sequentially inputting $|\alpha_{2}\rangle$ into modes $2$,$3$,...,$N$, and varying the phase over at least $2 \pi$, all phases of matrix $M$ can be directly determined. (b) Experimental realization. The device-under-test is a $3{\times}3$-mode fused-fiber beam splitter (FBS), which constitutes a $6{\times}6$ optical network including polarization. Orthogonal polarization modes are resolved using fibre polarization beam splitters (FPBS) at its outputs. Interferometric probe states between pair-wise input combinations are prepared with two polarization beam displacers (BD), and half-wave plates (HWP). The outputs are monitored with fast photo-diodes (PD) connected to an oscilloscope (OSC) while the phase $\phi$ is scanned.}
	\label{fig:setup}
\end{figure}

One way to achieve this, as shown in references~\cite{Peruzzo2011,laing2012sst}, is to insert two single photons into different input modes and to record nonclassical interference patterns between different combinations of output modes. Alternatively, such nonclassical interference can be simulated using two-mode coherent states with randomized relative phases~\cite{bromberg2009prl,keil2010pra}, but at the expense of an increased level of noise and reduced interference visibility.
Our method takes a more direct approach using a standard laser source split at a beam splitter, and with a varying relative phase between the resulting dual-mode coherent state, see Fig.~\ref{fig:setup}(a). This enables us to measure the nontrivial phases of the matrix $M$ directly without solving complex trigonometric equations, thus significantly simplifying the task of characterization. 
Our method works as follows: 
\begin{enumerate}
\item Send a coherent state with known intensity $I$ to input mode $j$, where other input modes are in the vacuum state, and measure the intensity $I_k$ from all output modes simultaneously. Using \eeqref{in-out-coh}, we obtain all the moduli
\begin{equation}
r_{jk}=\sqrt{\frac{I_k}{I}} , \ \ \ \ k=1,2,\dots, N .
\label{mean1}
\end{equation}
\item Send a coherent state $\ket{\alpha}$ to a $50{:}50$ beam splitter and use a phase shifter to control the relative phase between the output states $\ket{\alpha_1}$ and $\ket{\alpha_2}$ with the same intensity $I$, see Fig.~\ref{fig:setup}(a). 
Where $\ket{\alpha_1}$ goes to input mode `1' and $\ket{\alpha_2}{=}\ket{e^{i\phi}\alpha_1}$ to input mode $j$. The intensities of the output coherent states are given by
\begin{equation}
I_k = I\left| M_{1k} {+} M_{jk}e^{i\phi} \right|^2 .
\end{equation} 
As all elements in the first row and the first column are real, the above equation becomes, for $k{=}1$, 
\begin{align}
I_1 = I \left[ r_{11}^2 +  r_{j1}^2 + 2 r_{11} r_{j1} \cos(\phi) \right],
\end{align}
and, for $k{\neq}1$,
\begin{equation}
I_k = I \left[r_{1k}^2\! +  r_{jk}^2\! + 2 r_{1k} r_{jk} \cos(\phi+\theta_{jk})\right].
\label{mean-two-coh}
\end{equation}
When $I_1$ attains its maximum value we have $\phi{=}0$, and $\phi{=}\pi$ for its minimum value. This serves as our reference mode, and without loss of generality we always choose $I_1$ at its maximum such that $\phi{=}0$. Knowing this, we further sweep $\phi$ until $I_k$ attains its maximum value and using \eeqref{mean-two-coh} the unknown phases can be found as 
\begin{equation}
\theta_{jk}=2\pi{-}\phi \ .
\label{eq:phases}
\end{equation}
Repeating this procedure for $\ket{\alpha_2}$ input into mode $j{=}2,3,\ldots, N$ yields all the nontrivial phases $\theta_{jk}$ of the matrix $M$.
\end{enumerate}
\section{Experiment}

The network we characterize here is composed of one $3{\times}3$ non-polarizing fused fiber-optic beam splitter (FBS) with three $2{\times}2$ polarizing beam splitters at each of its output modes, as shown in Fig.~\ref{fig:setup}(b). By mapping onto orthogonal polarizations at the input of the initial FBS the whole network is described by a $6{\times}6$ matrix $M$. The input modes are labeled $\{1,2,\hdots,6\}{=}\{\ket{H}_1,\ket{V}_1,\ket{H}_2,\hdots,\ket{V}_3\}$, where $\ket{H}_1$ is the horizontally polarized mode for spatial mode `1' of the FBS.

\begin{figure}[b]
\centering
	\includegraphics[width=1\columnwidth]{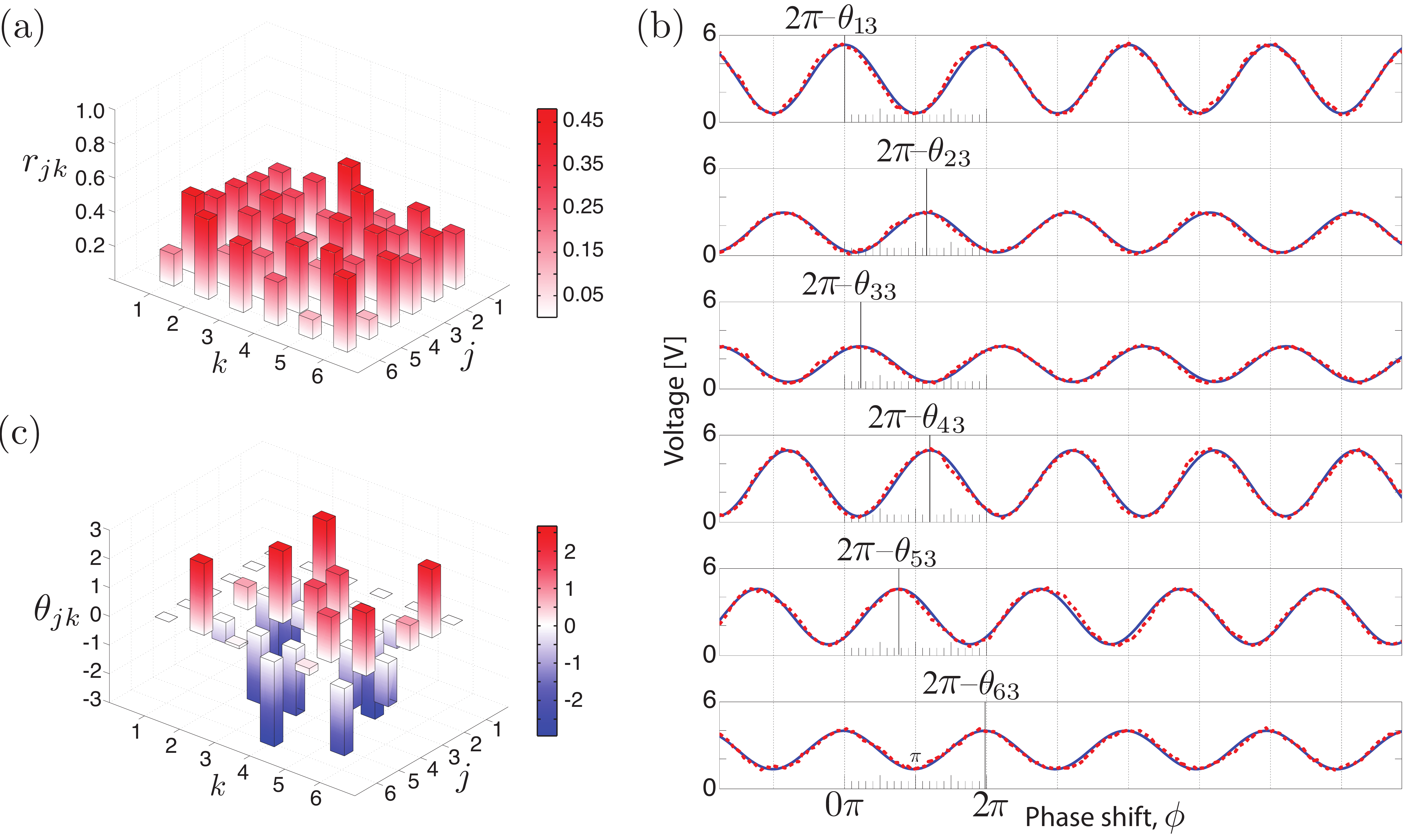}
	\caption{Experimental characterization of a linear-optical device. (a) Moduli $r_{jk}$ of the experimentally measured $M$. The x and y axes correspond to the input and output modes, $j$ and $k$ respectively. (b) Representative experimental data for obtaining $\theta_{jk}$ when injecting the variable-phase dual-mode coherent state into input modes $1$ and $3$. The amplitudes (voltage at output photo-diodes) of the six output modes (1-6 from top-to-bottom) oscillate as the phase $\phi$ is swept in time. Red and blue lines are measured data and theoretical fits to $A\cos(\phi-\theta_{jk})$ respectively. (c) Phases, $\theta_{jk}$, of the measured matrix $M$. The entire characterization method was performed $10$ times to obtain experimental uncertainties; error bars are not visible on the scales shown.}
	\label{fig:sincurves}
\end{figure}

In the setup we used a series of polarization beam displacers and half-wave plates to prepare input probe states, allowing for phase-stable interferometric measurements and polarization control for the input of the $N{\times}N$ network. The phase $\phi$ was controlled by a motorized linear micro-translation stage at input mode `1' to introduce an optical path difference of $0.1$~mm, at a speed of $0.05$~mm/s, between two inputs. Scanning over this short time window limits the effect of thermal drift on the classical interferometer, therefore removing the need for active phase locking. The outputs were coupled to fast photodiodes and monitored simultaneously on an oscilloscope while the phase $\phi$ was scanned. All characterization measurements were performed with a $100$~$\mu$W laser diode spectrally filtered to have a center wavelength of $820$~nm and a full-width half-maximum bandwidth of $2$~nm; thus the probe light had a coherence length of ${\sim}150$~$\mu$m, much shorter than the length of the optical circuit ${\sim}2$m. 

We first measured the 36 output intensities for the six individual inputs shown in Fig.~\ref{fig:sincurves}(a). We then recorded interference fringes for the pair-wise input combinations discussed above, and fitted sinusoidal curves to the resulting photocurrents to obtain experimental values for $\theta_{jk}$, see Figs.~\ref{fig:sincurves}(b) and~\ref{fig:sincurves}(c). From these measurements, we reconstructed the $6{\times}6$ matrix $M$.

\begin{figure}[b]
\centering
	\includegraphics[width=0.8\textwidth]{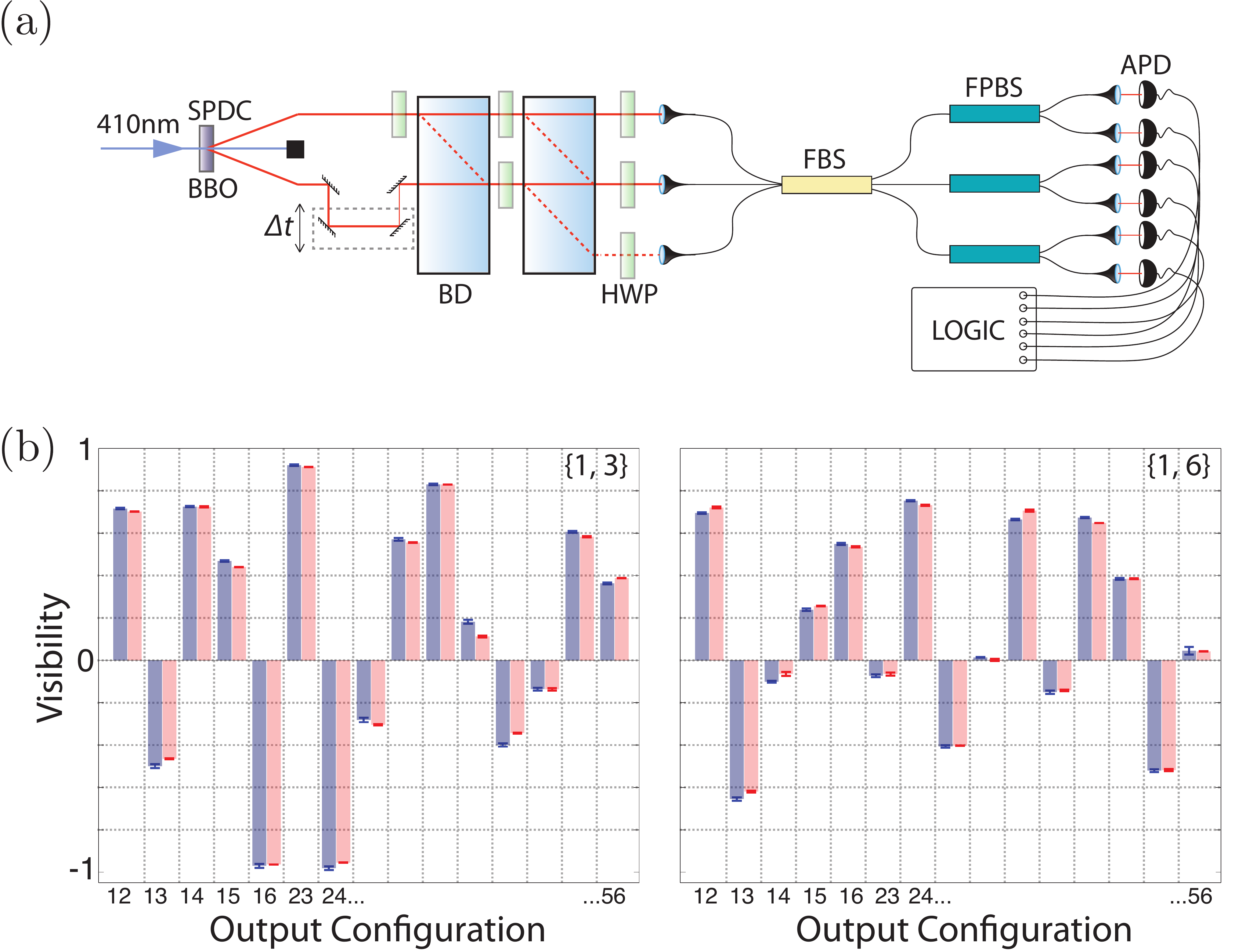}
	\caption{
Independent verification of the measured matrix $M$. 
	(a) Experimental schematic. A pair of $820$~nm photons is generated via type-I spontaneous parametric downconversion (SPDC) in a nonlinear $\beta$-barium-borate (BBO) crystal pumped with a mode-locked pulsed laser at $410$~nm. After being spectrally filtered  (FWHM~$2$~nm) individual downconverted photons are steered into the optical modes of the linear-optical network by a series of beam displacers (BD) and half-wave plates (HWP). The temporal overlap, $\Delta t$, between the input photons is controlled via a micro-translation stage at one of the inputs. Output photons are detected using avalanche photo-diodes (APD) whose coincident signals---monitored using a commercially available counting logic---are used to post-select two single photon events.
	(b) Measured nonclassical visibilities vs. predicted visibilities for photons input into modes $\{1, 3\}$ and $\{1, 6\}$. Red bars show the directly measured nonclassical visibilities; blue bars show the predictions from the measured matrix $M$; errors are given at the top of each data point. Numbers on the x-axis show the corresponding output modes.
}
	\label{fig:HOMS}
\end{figure}

We verify our experimentally obtained matrix $M$ by measuring two-photon interference inside the linear-optical network~\cite{PhysRevLett.59.2044}. As this is a fourth-order interference effect, it provides a suitable independent verification for the validity of $M$ that we obtained with our second-order interference method. We created and sent two single-photons into the $6{\times}6$ network, and measured the coincidences at all fifteen (6 choose 2) pairwise combinations of output modes; see Fig.~\ref{fig:HOMS}(a). 

The probability that two photons input to modes $i$ and $j$ simultaneously arrive at output modes $k$ and $l$ can be determined from the characterization matrix $M$. In the case of indistinguishable input photons this probability is given by
\begin{equation}
Q^{kl}_{ij}=\frac{1}{1+\delta_{ij}}\left|M_{ik}M_{jl}+M_{il}M_{jk}\right|^2,
\end{equation}
where $\delta_{jk}$ is Kronecker's delta function~\cite{mattle1995,Peruzzo2011}. Whereas in the case that the input photons are entirely distinguishable this probability is given by
\begin{equation}
C^{kl}_{ij}=\left|M_{ik}M_{jl}\right|^2+\left|M_{il}M_{jk}\right|^2.
\end{equation}
We can therefore determine the nonclassical interference visibility as $\mathcal{V}^{kl}_{ij}{=}(C^{kl}_{ij}{-}Q^{kl}_{ij})/C^{kl}_{ij}$. Experimentally $C$ and $Q$ are given by the coincidence count rates of photon pairs at the outputs $k$ and $l$ when there is a maximum temporal overlap between input photons (indistinguishable) and no overlap (distinguishable) respectively~\cite{PhysRevLett.59.2044}. We vary the temporal overlap between the single photon wave packets using an electronically controlled micro-translation stage on one of the input photons. The results for two different input configurations are shown in Fig.~\ref{fig:HOMS}(b). The obtained interference visibilities are in excellent agreement with those predicted by the experimentally measured matrix $M$. Higher-order photon terms from SPDC are believed to be responsible for the cases where the measured and predicted visibilities do not overlap within error. 


\section{Lossy networks}

\begin{figure}[t]
\centering
	\includegraphics[width=0.8\columnwidth]{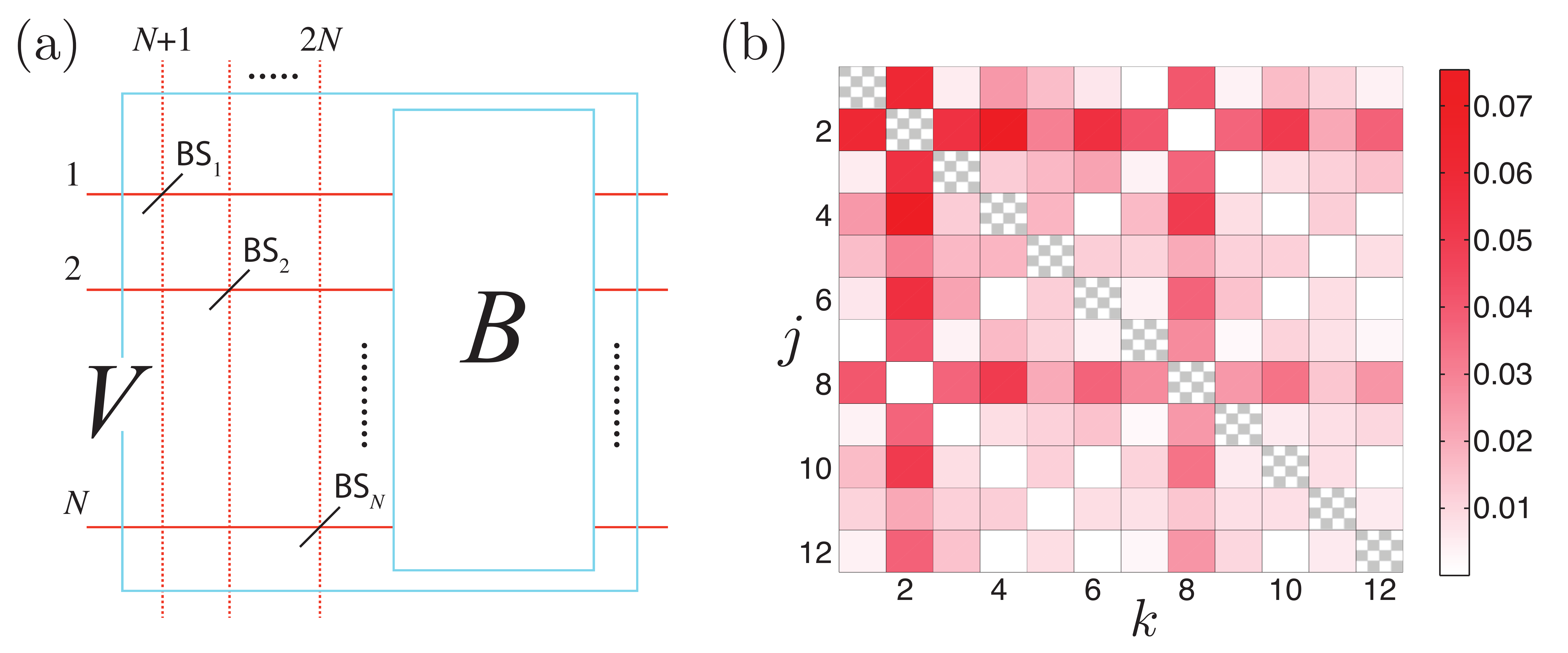}
	\caption{(a) Scheme for characterizing path-independent loss in the linear optical network. Virtual beam splitters are placed at each input mode of the optical network represented by a lossless matrix $B$. The resulting matrix $V$ is a $2N{\times}2N$ matrix and accounts for path-independent loss. (b) The matrix $VV^{\dagger}$. The diagonal hatched squares are equal to unity by construction and the off-diagonal elements are coloured according to their value given by the color bar.
}
	\label{fig:results}
\end{figure}

We now discuss how $M$ can be embedded into a matrix that is closer to unitarity. In principle, neglecting measurement error, there always exists a larger unitary matrix which fully accounts for all loss modes; however, it is not clear how to find this matrix for any arbitrary network. 
We show that if the optical loss is equal for different paths connecting specific inputs to outputs (path-independent loss), then an $N{\times}N$ network described by $M$ can be extended by $N$ virtual input and $N$ virtual output modes into a $2N{\times}2N$ network described by unitary matrix $V$, so $M$ will be a submatrix of $V$, see Fig.~\ref{fig:results}(a).   
The $2N{\times}2N$ network is obtained by adding $N$ beam splitters to each input of an $N{\times}N$ lossless linear network described by unitary matrix $B$. Each beam splitter is described by a $2{\times}2$ matrix
\begin{equation}
S_j=
\begin{pmatrix}[c]
\eta_j & -\sqrt{1-\eta_j^2} \\
\sqrt{1-\eta_j^2} &  \eta_j\\
\end{pmatrix}\ ,
\end{equation}
where $\eta_j$ ($0\leq\eta_j\leq 1$) is the transmissivity of the beam splitter.
Thus the total matrix describing $N$ beam splitters, by appropriately labeling input and output modes, is given by
\begin{equation}
S_{\rm tot}=
\begin{pmatrix}[c]
\bm{\eta} & -\tilde{\bm{\eta}} \\
\tilde{\bm{\eta}} &  \bm{\eta}\\
\end{pmatrix}\ ,
\end{equation}
where $\bm{\eta}$ and $\tilde{\bm{\eta}}$ are $N{\times}N$ diagonal matrices:
\begin{align}
\bm{\eta}&=\text{diag}\left(\eta_1,\eta_2, \dots, \eta_N \right),\nonumber \\
\tilde{\bm{\eta}}&=\text{diag}\left(\sqrt{1-\eta_1^2},\sqrt{1-\eta_2^2}, \dots, \sqrt{1-\eta_N^2} \right). \nonumber
\end{align}

The matrix $V$ describing the resulting network is obtained by multiplying the matrix describing $N$ parallel beam splitters by the matrix describing the lossless network that only acts on $N$ modes,
\begin{equation}
V=S_{\rm tot}\times 
\begin{pmatrix}[c]
 B & O \\
O &  \mathcal{I}\\
\end{pmatrix}\ =
\begin{pmatrix}[c]
\bm{\eta} B & -\tilde{\bm{\eta}} \\
\tilde{\bm{\eta}} B &  \bm{\eta}\\
\end{pmatrix}\ ,
\end{equation} 
where $O$ and $\mathcal{I}$ are $N{\times}N$ null and identity matrices, respectively. 
Using our method we measure the first block of $V$, $M{=}\bm{\eta} B$, and $\eta_j$ are obtained by measuring the output intensities $I_k$ when only a coherent state with intensity $I$ is sent to input $j$
\begin{equation}
\eta_j= \sqrt{\frac{1}{I}\sum_{k=1}^{N}I_k} \ .
\label{mean2}
\end{equation}
Thus the matrix $V$ can be experimentally determined.

In our case the $12{\times}12$ matrix $V$ is not unitary, since the off-diagonal elements in $VV^{\dagger}$, shown in Fig.~\ref{fig:results}(b), are non-zero albeit very small. This is indicative of both inevitable experimental uncertainty in the measured matrix elements and the effect of path-dependent loss in the network. Nevertheless, by comparing how close $VV^{\dagger}$ and $MM^{\dagger}$ are to identity in trace norm, one can see that $V$ more closely approximates a unitary matrix.

The question remains as to what unitary best describes the larger linear-optical network. It has been shown that the closest unitary matrix to $V$ can be found by using the polar decomposition $U{=}(VV^{\dagger})^{-\frac{1}{2}}V$~\cite{Fan:unitary}. The resulting unitary matrix $U$ does not noticeably alter the predicted two-photon interference visibilities of Fig.~\ref{fig:HOMS}(b), and therefore describes the larger network that contains our device
with a good approximation.


\section{Conclusion}

As photonic quantum technologies mature beyond small-scale demonstrations \cite{Lanyon:shors,lanyon2010tqc,Schreiber06042012}, there is an increasing requirement for methods of process validation and verification. Areas of direct applicability include the experimental characterization of waveguide arrays for quantum walks \cite{peruzzo2010qwc,sansoni2012tpb}, especially in three dimensions where current top-down imaging methods are not possible~\cite{owens2011two}. One of the most exciting applications pertains to the intermediate model of quantum computing---\textsc{BosonSampling}~\cite{aaronson2011ccl,Broome20122012,Metcalf20122012}. Our method provides an efficient means for characterizing large linear optical networks to obtain the scattering probabilities of multi-photon processes~\cite{Metcalf:2013fk}, which is a crucial component of \textsc{BosonSampling} experiment and was used recently in~\cite{Broome20122012}. For these larger characterizations an optical switch board built into wave-guide circuits~\cite{Sharkawy:02} could be used as opposed to a larger bulk-optics setup. A simpler way to achieve the required small displacements is to tilt optical components in the probe beam path~\cite{keil2010pra}.


\section*{Acknowledgments}

We thank Devon Biggerstaff, Shahla Nikbakht, and Peter Rohde for discussions. This work was supported in part by: Centre for Quantum Computation and Communication Technology (CE110001027), Centre for Engineered Quantum Systems (CE110001013), the Australian Research Council's Federation Fellow program (FF0668810), and the SFB program W1210-2 (Vienna Doctoral Program on Complex Quantum Systems - CoQuS) of the Austrian Science Fund (FWF); and the University of Queensland Vice-Chancellor's Senior Research Fellowship program. AF is supported by an Australian Research Council Discovery Early Research Award (DE130100240).

\end{document}